\documentstyle[11pt]{article}

\advance \topmargin by -\headheight
\advance \topmargin by -\headsep
     
\evensidemargin \oddsidemargin
\begin{document}

\renewcommand{\theequation}{\arabic{equation}}

\topmargin -.6in
\def\nonu{\nonumber}
\def\rf#1{(\ref{eq:#1})}
\def\lab#1{\label{eq:#1}} 
\def\br{\begin{eqnarray}}
\def\er{\end{eqnarray}}
\def\be{\begin{equation}}
\def\ee{\end{equation}}
\def\0{\nonumber}
\def\lb{\lbrack}
\def\rb{\rbrack}
\def\({\left(}
\def\){\right)}
\def\v{\vert}
\def\bv{\bigm\vert}
\def\lskip{\vskip\baselineskip\vskip-\parskip\noindent}
\relax
\newcommand{\nit}{\noindent}
\newcommand{\ct}[1]{\cite{#1}}
\newcommand{\bi}[1]{\bibitem{#1}}
\def\a{\alpha}
\def\b{\beta}
\def\ca{{\cal A}}
\def\cm{{\cal M}}
\def\cn{{\cal N}}
\def\cf{{\cal F}}
\def\d{\delta}
\def\D{\Delta}
\def\eps{\epsilon}
\def\g{\gamma}
\def\G{\Gamma}
\def\grad{\nabla}
\def\h{ {1\over 2}  }
\def\hc{\hat{c}}
\def\hd{\hat{d}}
\def\hg{\hat{g}}
\def\hp{ {+{1\over 2}}  }
\def\hm{ {-{1\over 2}}  }
\def\k{\kappa}
\def\l{\lambda}
\def\L{\Lambda}
\def\lg{\langle}
\def\m{\mu}
\def\n{\nu}
\def\o{\over}
\def\om{\omega}
\def\O{\Omega}
\def\p{\phi}
\def\pa{\partial}
\def\pr{\prime}
\def\ra{\rightarrow}
\def\rh{\rho}
\def\rg{\rangle}
\def\s{\sigma}
\def\t{\tau}
\def\th{\theta}
\def\ti{\tilde}
\def\wti{\widetilde}
\def\inte{\int dx }
\def\xb{\bar{x}}
\def\yb{\bar{y}}

\def\tr{\mathop{\rm tr}}
\def\Tr{\mathop{\rm Tr}}
\def\partder#1#2{{\partial #1\over\partial #2}}
\def\ds{{\cal D}_s}
\def\wtwo{{\wti W}_2}
\def\lie{{\cal G}}
\def\alie{{\widehat \lie}}
\def\dlie{{\cal G}^{\ast}}
\def\elie{{\widetilde \lie}}
\def\edlie{{\elie}^{\ast}}
\def\hlie{{\cal H}}
\def\wlie{{\widetilde \lie}}

\def\rlx{\relax\leavevmode}
\def\inbar{\vrule height1.5ex width.4pt depth0pt}
\def\IZ{\rlx\hbox{\sf Z\kern-.4em Z}}
\def\IR{\rlx\hbox{\rm I\kern-.18em R}}
\def\IC{\rlx\hbox{\,$\inbar\kern-.3em{\rm C}$}}
\def\one{\hbox{{1}\kern-.25em\hbox{l}}}

\def\PRL#1#2#3{{\sl Phys. Rev. Lett.} {\bf#1} (#2) #3}
\def\NPB#1#2#3{{\sl Nucl. Phys.} {\bf B#1} (#2) #3}
\def\NPBFS#1#2#3#4{{\sl Nucl. Phys.} {\bf B#2} [FS#1] (#3) #4}
\def\CMP#1#2#3{{\sl Commun. Math. Phys.} {\bf #1} (#2) #3}
\def\PRD#1#2#3{{\sl Phys. Rev.} {\bf D#1} (#2) #3}
\def\PRB#1#2#3{{\sl Phys. Rev.} {\bf B#1} (#2) #3}

\def\PLA#1#2#3{{\sl Phys. Lett.} {\bf #1A} (#2) #3}
\def\PLB#1#2#3{{\sl Phys. Lett.} {\bf #1B} (#2) #3}
\def\JMP#1#2#3{{\sl J. Math. Phys.} {\bf #1} (#2) #3}
\def\PTP#1#2#3{{\sl Prog. Theor. Phys.} {\bf #1} (#2) #3}
\def\SPTP#1#2#3{{\sl Suppl. Prog. Theor. Phys.} {\bf #1} (#2) #3}
\def\AoP#1#2#3{{\sl Ann. of Phys.} {\bf #1} (#2) #3}
\def\PNAS#1#2#3{{\sl Proc. Natl. Acad. Sci. USA} {\bf #1} (#2) #3}
\def\RMP#1#2#3{{\sl Rev. Mod. Phys.} {\bf #1} (#2) #3}
\def\PR#1#2#3{{\sl Phys. Reports} {\bf #1} (#2) #3}
\def\AoM#1#2#3{{\sl Ann. of Math.} {\bf #1} (#2) #3}
\def\UMN#1#2#3{{\sl Usp. Mat. Nauk} {\bf #1} (#2) #3}
\def\FAP#1#2#3{{\sl Funkt. Anal. Prilozheniya} {\bf #1} (#2) #3}
\def\FAaIA#1#2#3{{\sl Functional Analysis and Its Application} {\bf #1} (#2)
#3}
\def\BAMS#1#2#3{{\sl Bull. Am. Math. Soc.} {\bf #1} (#2) #3}
\def\TAMS#1#2#3{{\sl Trans. Am. Math. Soc.} {\bf #1} (#2) #3}
\def\InvM#1#2#3{{\sl Invent. Math.} {\bf #1} (#2) #3}
\def\LMP#1#2#3{{\sl Letters in Math. Phys.} {\bf #1} (#2) #3}
\def\IJMPA#1#2#3{{\sl Int. J. Mod. Phys.} {\bf A#1} (#2) #3}
\def\AdM#1#2#3{{\sl Advances in Math.} {\bf #1} (#2) #3}
\def\RMaP#1#2#3{{\sl Reports on Math. Phys.} {\bf #1} (#2) #3}
\def\IJM#1#2#3{{\sl Ill. J. Math.} {\bf #1} (#2) #3}
\def\APP#1#2#3{{\sl Acta Phys. Polon.} {\bf #1} (#2) #3}
\def\TMP#1#2#3{{\sl Theor. Mat. Phys.} {\bf #1} (#2) #3}
\def\JPA#1#2#3{{\sl J. Physics} {\bf A#1} (#2) #3}
\def\JSM#1#2#3{{\sl J. Soviet Math.} {\bf #1} (#2) #3}
\def\MPLA#1#2#3{{\sl Mod. Phys. Lett.} {\bf A#1} (#2) #3}
\def\JETP#1#2#3{{\sl Sov. Phys. JETP} {\bf #1} (#2) #3}
\def\JETPL#1#2#3{{\sl  Sov. Phys. JETP Lett.} {\bf #1} (#2) #3}
\def\PHSA#1#2#3{{\sl Physica} {\bf A#1} (#2) #3}
\def\PHSD#1#2#3{{\sl Physica} {\bf D#1} (#2) #3}

\newcommand\twomat[4]{\left(\begin{array}{cc}  
{#1} & {#2} \\ {#3} & {#4} \end{array} \right)}
\newcommand\twocol[2]{\left(\begin{array}{cc}  
{#1} \\ {#2} \end{array} \right)}
\newcommand\twovec[2]{\left(\begin{array}{cc}  
{#1} & {#2} \end{array} \right)}

\newcommand\threemat[9]{\left(\begin{array}{ccc}  
{#1} & {#2} & {#3}\\ {#4} & {#5} & {#6}\\ {#7} & {#8} & {#9} \end{array} \right)}
\newcommand\threecol[3]{\left(\begin{array}{ccc}  
{#1} \\ {#2} \\ {#3}\end{array} \right)}
\newcommand\threevec[3]{\left(\begin{array}{ccc}  
{#1} & {#2} & {#3}\end{array} \right)}

\newcommand\fourcol[4]{\left(\begin{array}{cccc}  
{#1} \\ {#2} \\ {#3} \\ {#4} \end{array} \right)}
\newcommand\fourvec[4]{\left(\begin{array}{cccc}  
{#1} & {#2} & {#3} & {#4} \end{array} \right)}

\begin{titlepage}
\vspace*{-2 cm}
\noindent
\begin{flushright}

\end{flushright}

\vskip 1 cm
\begin{center}
{\Large\bf Massive 3D Gravity Big-Bounce  } \vglue 1  true cm
{ H.L.C.Louzada}$^{*}$,
 { U.Camara dS}$^{*}$\footnote {e-mail: ulyssescamara@gmail.com} and { G.M.Sotkov}$^{*}$\footnote {e-mail: sotkov@cce.ufes.br, gsotkov@yahoo.com.br}\\

\vspace{1 cm}

${}^*\;${\footnotesize Departamento de F\'\i sica - CCE\\
Universidade Federal do Espirito Santo\\
29075-900, Vitoria - ES, Brazil}\\

\vspace{5 cm}

\end{center}

\normalsize
\vskip 0.5cm

\begin{center}
{ {\bf ABSTRACT}}\\
\end{center}
\noindent
The properties of an extension of the New Massive $3D$ Gravity by scalar matter with Higgs-like self-interaction are
investigated. Its peturbative unitarity consistency is verified for a family of cosmological Bounce solutions found by  the superpotential method. They correspond 
to the lower bound ${\lambda = -1}$ of the BHT unitarity window and describe eternally accelerated $3D$ Universe between two initial/final stable $dS_3$ vacua states. 
\end{titlepage}

The perturbative quantization of the Einstein gravity is known to introduce at  one-loop order in  the gravitational coupling $\kappa^2=16\pi G$ ``quadratic'' counterterms ${\mathcal{L}}_{ct} = \sqrt{-g}\Big(\alpha R^2+\beta R^{\mu\nu}R_{\mu\nu}+\ldots\Big)$ which cure its UV divergencies but violate unitarity (for $d\ge4$) \cite{velt}, \cite{stel}. In three dimensions the problem of perturbative consistency of quantum gravity model including specific ``high derivatives'' terms:
\begin{eqnarray}
S_{NMG}^{eff}(g_{\mu\nu},{\lambda_0},\Lambda)=\frac{1}{\kappa^2}\int dx^3\sqrt{-g}\Big[\epsilon R+64{\lambda_0} K-2\Lambda\Big] &,& K = R_{\mu\nu}R^{\mu\nu}-\frac{3}{8}R^2\label{acao}
\end{eqnarray}
called New Massive Gravity (NMG) was recently reexamined by Bergshoeff, Hohm and Townsend (BHT) \cite{BHT1}, \cite{BHT2}. It turns out that it is super-renormalizable and unitary (ghost free) under certain restrictions on the values of the cosmological constant $\Lambda$ and of the new scale (mass) parameter $\lambda_{0}$ for the both choices $\epsilon=\pm1$ of the ``right'' or ``wrong'' sign of the $R$-term \cite{BHT2}, \cite{oda}, \cite{liu}. The BHT - model (\ref{acao}) has two propagating degrees of freedom and unlike $3D$ Einstein gravity ($i.e.$ $\lambda_{0}=0$ case) it admits interesting (vacuum) solutions - black holes, gravitational waves, etc. \cite{clem}, \cite{wave}. Its Newtonian limit for $\epsilon=-1$ reproduces an attractive gravitational force \cite{tekin}. 

The New Massive Gravity (\ref{acao}) can be realized perturbatively as a
massive spin two quantum field $\hat{h}_{\mu\nu}$ ($\hat{g}_{\mu\nu}=\bar{g}_{\mu\nu}+\kappa \hat{h}_{\mu\nu}$) - theory with unusual kinetic term and specific self-interactions in external constant curvature background $\bar{g}_{\mu\nu}$.
Such  a procedure provides well defined rules for studying the quantum graviton ($\hat{h}_{\mu\nu}$) properties and eventually it might  shed some light on the nature of the ``quantum $3D$ geometry'' behind it. Thus we have an example of consistent (3D) quantum gravity model that shares many of the (desired) properties of $4D$ gravity. It is therefore natural to address the question about the perturbative quantum consistency of the New Massive Gravity  coupled to matter fields. Since we are interested to study the Cosmological consequences  of such BHT-matter model, the simplest and most representative candidate of matter QFT is the one of scalar field $\sigma(x^{\mu})$ with Higgs - like self-interaction:
\begin{eqnarray}
{\mathcal{L}}_{matt}(\sigma,g_{\mu\nu}\vert\gamma,m_{\sigma})=\sqrt{-g}\Big(-\frac{1}{2}\partial_{\mu}\sigma\partial^{\mu}\sigma-V(\sigma)\Big) &,&
V(\sigma)=\frac{\gamma}{4}\Big(\sigma^2-\frac{m_{\sigma}^{2}}{2\gamma}\Big)^2-2\Lambda\label{lagrange}
\end{eqnarray}
The parameters $\gamma$, $m_{\sigma}$ and $\Lambda$ above denote the renormalized coupling constant $\gamma$, ``Higgs mass'' $m_{\sigma}$ and the vacuum energy $\Lambda$ and they are assumed to depend on the ``graviton mass" scale $\lambda_{0}=\frac{1}{64m^2}$. Since in four dimensions (in curved space) this model is known to be renormalizable \cite{BD}, \cite{bunch} one  expects that it  is power - counting super-renormalizable in three dimensions. Note that the relevant (lowest order in $\gamma$) counter-terms to ${\mathcal{L}}_{matt}$ are again in the form $\tilde{\alpha}R^{2}+\tilde{\beta}R_{\mu\nu}R^{\mu\nu}$  \cite{BD}, $i.e.$ they  do coincide with the gravitational ones (in the case of minimal coupling $\xi=0$).
We further assume that the corresponding (vacuum) ``semiclassical'' equations  \cite{BD}, \cite{ford}:
\begin{eqnarray}
\bar{R}_{\mu\nu}-\frac{1}{2}\bar{g}_{\mu\nu}\bar{R}+\Lambda\bar{g}_{\mu\nu} = \frac{1}{2}\kappa^{2}
<vac|\hat{T}_{\mu\nu}(h_{\mu\nu},\sigma)|vac>\label{quantum}
\end{eqnarray}
for this BHT - matter model (to lowest order in the both couplings $\kappa^2$ and $\gamma$) with the UV counter-terms added and when adiabatic vacuum states $|vac>=| g_{\mu\nu}^{vac},\sigma_{\pm}>$, ($V'(\sigma_{\pm})=0$) exist, are represented by the $\it {classical}$ equations obtained from the following ``effective" Matter-Gravity action:
\begin{eqnarray}
S_{NMGM}^{eff}=\frac{1}{\kappa^2}\int dx^3\sqrt{-g}\Big[\epsilon R+64\lambda_{0}K-\frac{1}{2}|\nabla\sigma|^2-V(\sigma)\Big]\label{512}
\end{eqnarray}
The BHT - cosmological constant $\Lambda$ is now included in the potential (\ref{lagrange}), $i.e.$ $V\big(\sigma^2_{\pm}=\frac{m_{\sigma}^2}{2\gamma}\big)=-2\Lambda$ for the case $\epsilon\lambda_{0}<0$. In order to describe in a compact form the corrections ($i.e.$ backreaction) to the classical background $\bar {g}_{\mu\nu}$
($\bar {R}_{\mu\nu} = R_{\mu\nu}(\bar {g}_{\mu\nu})$ etc.)  due to the quantum fluctuations  of both $\hat{h}_{\mu\nu}$ and $\hat{\sigma}$, as usually (see ref. \cite{ford}) all the higher derivatives  contributions  are included in the semiclassical ``source" in the r.h.s. of eqs.(\ref{quantum}).

The present paper is devoted to the investigation of the properties  of homogeneous and isotropic spatially flat $k=0$ 
($-,+,+$) $3D$ Universe :
\begin{eqnarray}
ds_{3}^{2}=-dt^2+a_{eff}^{2}(t)(dx^2+dy^2)\label{anzatz} 
\end{eqnarray} 
described by the matter extension (\ref{512}) of BHT-model (\ref{acao}).
Our ${\it{problem}}$ is therefore to find  analytic solutions of the corresponding $3D$ ``Massive Cosmological'' equations\footnote{in what follows we use the convention $c=\kappa=\hbar=1$} : 
\begin{eqnarray}
\ddot{\sigma}&+&\dot{\sigma}\dot{\varphi}+V'(\sigma)=0\nonumber\\
\ddot{\varphi}(1&+&8\epsilon\lambda_{0}\dot{\varphi}^2)+\frac{1}{2}\dot{\varphi}^2(1+4\epsilon\lambda_{0}\dot{\varphi}^2)+\epsilon\Big(\frac{1}{2}\dot{\sigma}^2-V(\sigma)\Big) =0\nonumber\\
\epsilon(\dot{\sigma}^2&+&2V(\sigma))-\dot{\varphi}^2(1+4\epsilon\lambda_{0}\dot{\varphi}^2)=0\label{sist}
\end{eqnarray}
derived from the  action (\ref{512}) above. We are further interested in studying specific features of these solutions as for example: the effects of the matter on the BHT - unitarity conditions (u.c.'s), the conditions of absence of initial/final singularities, how many different periods of acceleration and/or deceleration they describe etc.

Observe that eqs. (\ref{sist}) for the  scale factor $a(t)=e^{\frac{\varphi(t)}{2}}$ and for the scalar field $\sigma(t)$ are of second order although the action (\ref{512}) and  the eqs. (\ref{quantum}) include (up to) fourth order time derivitives of $g_{\mu\nu}$. The special BHT unitarity motivated choice of the relative coefficients $\beta=-\frac{3}{8}\alpha=64\lambda_{0}$ (that fix the form of the K-term for $D=3$) together with the FRW anzatz (\ref{anzatz}) are responsable for the cancellation of all the terms involving $\varphi^{(III)}$ and $\varphi^{(IV)}$ derivatives and they cure another disease of all ``higher derivatives'' gravities. The family of solutions we are considering does  have well defined initial values Cauchy problem ($i.e.$ ``causality'')  for {\it{all  the values}} of the parameters $\epsilon$, $\lambda_{0}$, $\gamma$, $m_{\sigma}^{2}$, $i.e.$ independently of whether or not the BHT u.c.'s \cite{BHT2} are satisfied. The construction of all the solutions of this second order system of nonlinear eqs. (\ref{sist}) is however rather difficult problem, although numerical methods are indeed available and largely used for this type of dynamical systems. One has to be able to further analyze how their properties depend on the initial conditions ($\sigma_{0}$, $\dot{\sigma}_{0}$, $\varphi_{0}$, $\dot{\varphi}_{0}$) and on the values of the parameters $\epsilon$, $\lambda_{0}$, $\gamma$, $m_{\sigma}$ as well.

At this place we take a different route known as ``superpotential'' method, which provides very special analytic (non-perturbative in $\gamma$ and $\lambda_{0}$) solutions of eqs. (\ref{sist}). It is widely used in the construction of domain walls (DW) solutions of different (super) gravity models in arbitrary dimensions (see for example refs. \cite{dw}, \cite{pt} for $\lambda_{0}=0$ case). It consists in the introduction of an auxiliary function $W(\sigma)$ (called superpotential) such that:
\begin{eqnarray}
V(\sigma)&=&2\epsilon W^2(1+16\epsilon\lambda_{0} W^2)-2(W')^2(1+32\epsilon\lambda_{0} W^2)^2\nonumber\\
\dot{\varphi}&=&-2\epsilon W(\sigma)\nonumber\\
\dot{\sigma}&=&2W'(\sigma)(1+32\epsilon\lambda_{0} W^2)\label{W}
\end{eqnarray}
For $\lambda_{0}\neq0$ it represents an adapted version of the Low-Zee superpotential \cite{zee} introduced in the context of DW's solutions of $d=5$ Gauss-Bonnet improved gravity. The statement is that for each function $W(\sigma)$ that reproduces our matter potential (\ref{lagrange}) the solutions of the first order system (\ref{W}) are solutions of the eqs. (\ref{sist}) as well.

We next consider a particular linear superpotential  $W(\sigma)=B\sigma$ , which leads to a family of $\sigma^{4}$ - like ($Z_{2}$ symmetric $\sigma\rightarrow-\sigma$) potentials (\ref{lagrange}) with 
\begin{eqnarray}
\gamma = 128\lambda_{0}B^4(1-64\lambda_{0}B^2) &,&
m_{\sigma}^2 = -8\epsilon B^2(1-64\lambda_{0}B^2)\nonumber\\
\sigma_{\pm}^{2} = -\frac{1}{32\epsilon\lambda_{0}B^2}, \ \Lambda=\frac{1}{64\lambda_{0}}=m^2 &,&
\bar{m}_{\sigma}^{2}=V''(0)=-\frac{1}{2}V''(\sigma_{\pm})=-\frac{m_{\sigma}^2}{2}\label{iff}
\end{eqnarray}
Hence depending on the $\epsilon\lambda_{0}$ sign (and whether $B^2< m^2$ or $B^2> m^2$ ) we have normal or inverted double-well potentials for $\epsilon\lambda_{0}<0$ or the standard $\sigma^4$ - potentials for $\epsilon\lambda_{0}>0$.
The fact that the form of the matter potentials (encoded here in $\gamma=\gamma(\lambda_{0})$ and $m_{\sigma}^2=m_{\sigma}^2(\lambda_{0})$) does depend on the gravitational mass scale $m^{2}=\frac{1}{64\lambda_{0}}$ is an intrinsic property of the superpotential method (see eqs. (\ref{W})) when applied for models which actions contain $K$ - type terms (\ref{acao}).

Note that  for the particular choice of $W(\sigma)$ we consider, the parameters $\{\gamma,m_{\sigma}^2,m^2\}$ are not independent. As one can see from eqs.(\ref{iff}), they describe a special surface 
\begin{eqnarray}
m_{\sigma}^{6} = 32\gamma m^2(m_{\sigma}^2+4\epsilon\gamma) &,&
1+\epsilon\frac{m_{\sigma}^2}{2m^2}>0\label{m6}
\end{eqnarray}
in the parameter space of the BHT - matter model (\ref{512}) . Taking the effective masses $m_{\sigma}^2$ and $m^2$ as independent variables we realize that  the  coupling constant $\gamma$ (or equivalently  the  ``superpotential" parameter $B^2$) are given by:
\begin{eqnarray}
\gamma_{\pm}=-\epsilon\frac{m_{\sigma}^2}{8}\Big(1\pm\sqrt{1+\epsilon\frac{m_{\sigma}^2}{2m^2}}\Big)\equiv-\epsilon\frac{m_{\sigma}^2B_{\pm}^2}{4m^2}\label{galinice2}
\end{eqnarray}
$i.e.$ they take two different values $\gamma_{\pm}$ (and $B_{\pm}^2$) when $\frac{m_{\sigma}^2}{\epsilon m^2}<0$ and only one $\gamma_{-}$ (or $B_{-}^2$) in the case $\frac{m_{\sigma}^2}{\epsilon m^2}>0$. Together with the signs of $m_{\sigma}^2$ and $m^2$ they characterize the particular shapes of the potential $V(\sigma\vert\gamma_{\pm},m_{\sigma}^2)$ for which we are able to solve analytically eqs. (\ref{sist}) by the superpotential method. We will consider mainly the case $\bar{m}_{\sigma}^2\gamma<0$ ($i.e.$ $\epsilon\lambda_{0}<0$) which describes:

(a) $\epsilon=-1$, $\lambda_{0}>0$ BHT - model with double-well (Higgs) potential $V_{-}(\sigma)$.

(b) $\epsilon=1$, $\lambda_{0}<0$ BHT - model with $\it{inverted}$ double well potential $V_{+}(\sigma)$. 

In both cases we have $\sigma_{\pm}^2>0$ and therefore $\sigma_{\pm}=\pm\sqrt{\frac{m_{\sigma}^2}{2\gamma}}$ are the minima ($\epsilon=-1$) or the maxima ($\epsilon=1$) of the potentials $V_{\pm}$. For these ``vacuum'' values $\sigma_{\pm}$ of the field $\sigma(t)$, the BHT - matter model (\ref{512})  reduces to the pure BHT - gravity (\ref{acao}) for a particular value of the cosmological constant\footnote{In the BHT notations \cite{BHT2} we have $\lambda_{BHT}=-\frac{\Lambda}{m^2}$ and therefore both $\epsilon\lambda_{0}<0$ cases correspond to the limiting value $\lambda_{BHT}=-1$.} $\Lambda=m^2=\frac{1}{64\lambda_{0}}$, $i.e.$ we have $\Lambda<0$ for $\lambda_{0}<0$ ($\epsilon=1$) and $\Lambda>0$ for $\lambda_{0}>0$ ($\epsilon=-1$). The effective cosmological constant $\Lambda_{eff}$ is defined by the values of $3D$ curvature $R^{(3)}$ on the corresponding vacuum solutions $\big(g_{\mu\nu}^{vac,\pm},\sigma_{\pm}\big)$:
\begin{eqnarray}
R^{(3)}(g_{\mu\nu}^{vac},\sigma_{\pm})=2\ddot{\varphi}+\frac{3}{2}\dot{\varphi}^2\equiv-8\epsilon(W')^2(1+32\lambda_{0}\epsilon W^2)+6W^2\label{R}
\end{eqnarray}
which gives $R^{(3)}(g_{\mu\nu}^{vac},\sigma_{\pm})=-\frac{3}{16\epsilon\lambda_{0}}\equiv6\Lambda_{eff}^{\pm}$. 
Therefore in both cases when $\epsilon\lambda_{0}<0$,  the $\Lambda_{eff}^{\pm}=2\vert m^2\vert>0$ is positive,
and the scale factor $a_{vac}^2(t)=e^{2\sqrt{\Lambda_{eff}}t}\equiv a_{dS_{3}}^2$ represents $dS_{3}$ geometry in agreement with the generic relation between $\Lambda_{eff}$ and $\Lambda$ \cite{BHT2}:
\begin{eqnarray}
\Lambda_{eff}^{\pm}= - 2m^2\Big(\epsilon\pm\sqrt{1+\lambda_{BHT}}\Big)\label{14}
\end{eqnarray}

We have two options for the initial conditions of $a(t)$ and $\sigma(t)$:
\begin{itemize}
\item \underline{type (1)}: $\sigma^2(t)\le\sigma_{\pm}^2, \ i.e.$
\begin{center}
$\sigma(t\rightarrow\pm\infty)=\sigma_{\pm}, \ \sigma_{-}\le\sigma(t)\le\sigma_{+}$

$a_{vac}^{\pm}=a_{dS_{3}}(t)=e^{\sqrt{\Lambda_{eff}}t}$
\end{center}
\item \underline{type (2)}: $\sigma^2(t)\ge\sigma_{\pm}^2$
\begin{itemize}
\item(2a) 
\begin{center}
$\sigma_{-}<\sigma(t)\le\sigma(t=t_{0})=\infty$

$a_{vac}^{-}=a_{dS_{3}}$ but $a(t_{0})=0$ and $R^{(3)}(t=t_{0})=\infty$
\end{center}
\item(2b)
\begin{center}
$\sigma_{+}\le\sigma(t)<\sigma(t=t_{0})=\infty$
\end{center}
\end{itemize}
\end{itemize}
Let us consider first the $\it{non-singular}$ solutions corresponding to the initial conditions of the {\it{type}} (1) above, $i.e.$ the ones interpolating between the two vacuum states: $\vert vac>=\vert\sigma_{\pm},dS_{3}(\Lambda_{eff})>$. We find by direct integration of eqs. (\ref{W})  two exact  solutions:
\begin{eqnarray}
\sigma_{\pm}(t\vert \epsilon)=\sqrt{\frac{m_{\sigma}^2}{2\gamma_{\pm}}}\tanh(b_{\pm}(t-t_{0})) &,&
a_{\pm}^{2}(t\vert\epsilon)=e^{\varphi_{\pm}(t\vert\epsilon)}=\Big(\cosh b_{\pm}(t-t_{0})\Big)^{\delta_{\pm}}\label{sol}
\end{eqnarray}
where $\delta_{\pm}$ and $b_{\pm}$ are given by:
\begin{eqnarray}
b_{\pm}=\frac{m^2}{\sqrt{2\vert m^2\vert}}\Big(1\pm\sqrt{1+\epsilon\frac{m_{\sigma}^2}{2m^2}}\Big)\ &,&
\delta_{\pm} = -\epsilon\frac{m_{\sigma}^2}{2\gamma_{\pm}}=-2\epsilon\frac{\vert m^{2}|}{B_{\pm}^2}\label{galinice4}
\end{eqnarray}
Observe that $\sigma_{\pm}(t)$ formally  coincide with the $ \it {flat}$  $\it {space}$ QM "instanton" solution\footnote{note that in our  curved space example  the time is not euclidean (see eq. (\ref{anzatz})) so that it just looks like an instanton} for the double well potential $V_{\pm}(\sigma)$, but now $\gamma$ and $m_{\sigma}^2$ are related by eq. (\ref{m6}). The scale factor $a_{\pm}^{2}(t)$ resembles  the 
well known bounce solution for closed $k=1$ $dS_{3}$ Universe ($\lambda_{0}=0$, $\sigma=const$) as one can see by comparing the behavior of the corresponding Hubble parameters:
\begin{eqnarray*}
H_{\pm}^{k=0}(t,\epsilon,\lambda_{0})=\frac{\dot{a}_{\pm}}{a_{\pm}}=-\epsilon B_{\pm}\sigma_{\pm}(t) &,& 
H^{k=1}(t)\sim\tanh 2\sqrt{\Lambda}(t-t_{0})
\end{eqnarray*}

The explicit form of the curvature scalar (\ref{R}), $i.e.$
\begin{eqnarray}
R_{\pm}(t)=\frac{8B_{\pm}^2}{\cosh^2 b_{\pm}(t-t_{0})}\Big(-\epsilon+\frac{3}{4}\sigma_{\pm}^{2}\sinh^2 b_{\pm}(t-t_{0})\Big)\label{galinice5}
\end{eqnarray}
confirms that the bounce solution (\ref{sol}) represents non-singular asymptotically $dS_{3}$ ($(a)dS_{3}$) geometry for all $t\in (-\infty,\infty)$, in particular
\begin{eqnarray}
R^{(3)}(t\rightarrow\pm\infty)=12\vert m^2\vert=6\Lambda_{eff}^{\pm}=\frac{6}{L_{gr}^2}, \ L_{gr}\gg l_{pl}
\end{eqnarray}
Another important feature of this $(a)dS_{3}$ Bounce solution (\ref{sol}) is encoded in the behavior of deceleration parameter:
\begin{eqnarray}
q_{\pm}(t)\equiv-1-2\frac{\ddot{\varphi}}{\dot{\varphi}^2}=-1-\frac{s_{\pm}(\epsilon)}{\sinh^2(b_{\pm}(t-t_{0}))}\label{q}
\end{eqnarray}
where $s_{\pm}$ are real numbers
\begin{eqnarray*}
s_{\pm}(\epsilon=-1)=\frac{1}{2}\Big(1\pm\sqrt{1-\frac{m_{\sigma}^2}{2m^2}}\Big)=\frac{B_{\pm}^2}{m^2} &,&
s_{-}(\epsilon=1)=\frac{1}{2}\Big(1-\sqrt{1+\frac{m_{\sigma}^2}{2m^2}}\Big)=-\frac{B_{-}^2}{\vert m^2\vert}
\end{eqnarray*}
They are determined by the  $ \it{ratio}$ of the two scales
\begin{eqnarray}
L_{gr}^2=\frac{1}{\vert\Lambda_{eff}\vert}=\frac{1}{2\vert m^2\vert} &,& \ L_{\sigma}^2=\frac{1}{\vert m_{\sigma}^2\vert}\nonumber\\ 
l_{pl}\ll L_{gr}&<& L_{\sigma}\label{20}
\end{eqnarray}
introduced by the graviton mass $M_{gr}^2=2\vert m^2\vert$ and by the scalar field mass\footnote{or by $\bar{m}_{\sigma}^2=-\frac{m_{\sigma}^2}{2}$ for the case $\epsilon=1$, since in this case $m_{\sigma}^2<0$, $i.e.$ it describes tachyons.} $\vert m_{\sigma}^2\vert$.

Note that our Bouncing $3D$ Universe is $\it{almost}$ invariant under time reflections $T:t\rightarrow2t_{0}-t$ ($t\rightarrow -t$ for $t_{0}=0$). The $\it{exact}$ invariance is achieved by combining $T$ with the $ Z_{2}$ - charge conjugation of $\sigma(t)$:
\begin{eqnarray*}
C: \  \ C\sigma_{(k)}(t)=-\sigma_{(k)}(t)=\sigma_{(ak)}(t)
\end{eqnarray*}
which interchanges ``kink'' $\sigma_{(k)}(t)$ with the ``anti-kink'' $\sigma_{(ak)}(t)=-\sigma_{(k)}(t)=\sigma_{(k)}(-t)$ ($i.e.$ $\sigma_{+}\rightarrow\sigma_{-}$ in the vacuum sector). Hence the past of the ``kink'' coincides with the future of the "anti-kink". This fact indeed reflects the CPT - invariance ($Px^{i}= -x^{i}$, $i=1,2$) of the BHT - matter model (\ref{512}). 

Although the forms of the Bounce solutions (\ref{sol}) in the two cases $\{\epsilon=-1,\lambda_{0}>0\}$ and $\{\epsilon=1,\lambda_{0}<0\}$ considered above are $\it {almost}$ identical\footnote{the only difference is in the scale factors $a^2(\epsilon=-1)=a^{-2}(\epsilon=1)$.}, the features of the Universe evolutions they describe are completely different. The origin of the qualitatively different physics behind is in the form of the matter interaction $V_{\pm}$: \emph{stable} vacuum double well potential for $\epsilon=-1$ versus \emph{unstable} ``maxima'' of the inverse double well potential for $\epsilon=1$ case \footnote{ intrinsically related  to the wrong/right signs $\epsilon=\pm1$ of the Einstein-Hilbert term $\epsilon R$ in the  action (\ref{512})}. Consider first the $\{\epsilon=-1,\lambda_{0}>0\}$ case. Since $s_{\pm}(\epsilon=-1)>0$ is always positive, the deceleration (\ref{q}) $q_{\pm}(t\vert\epsilon =-1)<0$ is \emph{negative} for all $t\in (-\infty,\infty)$ and therefore we have an \emph{eternally} accelerated Universe. It starts its evolution from well defined (stable) $dS_{3}$ vacuum state ($\sigma_{-}$, $dS_{3}(\Lambda_{eff})$) at the left minima of the matter potential living a period of (accelerated) expansion ($\dot{\varphi}>0$ for $B<0$) for all $t<t_{0}$. At the moment $t=t_{0}$ (say $t_{0}=0$) the Universe enters in the contraction ($\dot{\varphi}<0$) period towards the  new (``final'') vacuum state ($\sigma_{+}$, $dS_{3}(\Lambda_{eff}^{+})$) with $\Delta\sigma=\sigma_{+}-\sigma_{-}=m_{\sigma}\sqrt{\frac{2}{\gamma}}$ and $\Lambda_{eff}^{+}=\Lambda_{eff}^{-}=2m^2$. 

In the second case $\{\epsilon=1,\lambda_{0}<0\}$, that corresponds to the inverted double well potential $V_{-}(\sigma)$,  due to the negative values of $s_{-}(\epsilon=1)= -\frac{B_{-}^{2}}{\vert m^2\vert}$, the deceleration $q_{-}(t\vert\epsilon=1,\lambda_{0}<0)$ changes twice its sign. Therefore we have in this case two acceleration epochs (initial and final ones) and one period of deceleration. By studying the properties of scale factor $a^{2}(t)$ and its derivatives $\dot{a}$ and $\ddot{a}$ we can reconstruct the history of this non-singular Bounce - like $3D$ Universe. At the infinite past $t\rightarrow-\infty$ when: $\sigma(t)\rightarrow\sigma_{-}$, $a_{-}^{2}(t)\rightarrow e^{2\vert m\vert\sqrt{2}t}$ it starts from the unstable maximum ($\sigma_{-}$, $dS_{3}(\Lambda_{eff}^{-})$) of  $V_{-}(\sigma)$  its first (semi-infinite) period of acceleration and contraction with $R^{(3)}(t\vert\epsilon=1,\lambda_{0}<0)$ decreasing from $6\Lambda_{eff}^{-}=2\vert m^2\vert$ to $R^{(3)}(t_{1})=0$. At the moment $t_1<0$ 
\begin{eqnarray*}
t_{1}=-\frac{1}{b_{-}}\ln\Bigg(\sqrt{\frac{2\vert s_{-}\vert}{3}}+\sqrt{\frac{2\vert s_{-}\vert}{3}+1}\Bigg),
\end{eqnarray*}
the curvature $R^{(3)}(t_{1}<t<-t_{1})<0$ changes its sign entering in the epoch of negative curvature of finite duration $T_{-}=2\vert t_{1} \vert$ up to the moment $\bar{t}_{1}=-t_{1}>0$ when it changes its sign once more towards the finite \emph{unstable} $\{\sigma_{+},dS_{3}\}$ - state. Concerning the accelaration we observe that at the moment $\vert t_{a}\vert>\vert t_{1}\vert$:
\begin{eqnarray*}
\vert t_{a}\vert=\frac{1}{b_{-}}\ln\Big(\sqrt{\vert s_{-}\vert}+\sqrt{\vert s_{-}\vert+1}\Big)
\end{eqnarray*}
the Universe enters in deceleration period which takes place over a finite period of time $T_{dec}=2\vert t_{a}\vert$. The second (final) epoch of acceleration and expansion starts at $t=\vert t_{a}\vert$ towards its ``\emph{future eternity}'' $\{\sigma_{+},dS_{3}\}$ at $t\rightarrow\infty$.

An equivalent (but more compact) description of the Universe evolution is based on the so called equations of state: $p_{eff}=p_{eff}(\rho_{eff})$, where the effective pressure $p_{eff}$ and energy density $\rho_{eff}$ are defined by the diagonal r.h.s. of eqs. (\ref{quantum}):
\begin{eqnarray*}
< vac\vert T^{\mu}_{\nu}(\sigma,h_{\alpha\beta})\vert vac >=(-\rho_{eff},p_{eff},p_{eff})
\end{eqnarray*}
and it appears to be a simple consequence of the eqs. (\ref{sist}), (\ref{W}):
\begin{eqnarray}
p_{eff}(t)=8\epsilon m^2 s_{\pm}+(s_{\pm}-1)\rho_{eff}(t), \ \rho_{eff}>0\label{p1}
\end{eqnarray}
In the $\{\epsilon=-1,\lambda_{0}>0\}$ case we have                
\begin{eqnarray}
p_{eff}(t)=-8B_{\pm}^{2}-\rho_{eff}(t)\Big(1-\frac{B_{\pm}^2}{m^2}\Big) &,& \ B^2_{\pm}\le m^2\label{p2}
\end{eqnarray}
$i.e.$ $p_{eff}<0$ is always negative and that is why the $3D$ Universe is eternally accelerated. In the second case $\{\epsilon=1,\lambda_{0}<0\}$ the pressure changes twice its sign (at the moments $\pm t_{a}$), due to the existence of critical value of the energy density
\begin{eqnarray*}
\rho_{eff}^{cr}(\epsilon=1,\lambda_{0}<0)=\frac{8B_{-}^2\vert m^2\vert}{\vert m^2\vert+B_{-}^2}
\end{eqnarray*}
where $p_{eff}$ vanishes. This fact leads to the observation that there exist two periods of negative pressure  and another one  of positive pressure  between them.

An important problem in the classical (and quantum) Cosmology concerns the initial and/or final conditions (i/f. c.'s). Namely, whether and how the particular features of the Universe evolution depend on the corresponding i/f. c.'s. The BHT - matter model (\ref{512}) provides an example of two distinct families of such conditions (type (1) and (2) above) that lead to qualitatively different Universe histories: \emph{non-singular} Big-Bounce for the type (1) ($\sigma^2\le\sigma_{\pm}^2$) and the $\it {singular}$ Big - Bang for the type (2) ($\sigma^2\ge\sigma_{\pm}^2$) for the \emph{same values} of all the parameters $\{\kappa^2,\lambda_{0},m_{\sigma}^2\}$ of the model. The corresponding \emph{type} (2) solutions:
\begin{eqnarray}
\tilde{\sigma}(t\vert\epsilon)= \sqrt{\frac{m_{\sigma}^2}{2\gamma_{\pm}}}\coth\big(b_{\pm}(t-t_{0})\big) &,&
\tilde{a}_{\pm}^2(t\vert\epsilon) = \vert\sinh b_{\pm}(t-t_{0})\vert^{\delta_{\pm}}\label{23}
\end{eqnarray}
are ``singular'' at $t=t_{0}$, $i.e.$
\begin{eqnarray}
\tilde{\sigma}(t\rightarrow t_{0})\rightarrow\infty, \ \dot{\tilde{\varphi}}(t\rightarrow t_{0})\rightarrow\infty, \ \tilde{R}^{(3)}(t\rightarrow t_{0})\rightarrow\infty\label{24}  
\end{eqnarray}
and they exist in two disconnected periods of time. For $t\in (-\infty,t_{0})$ we have the same vacuum $\{\sigma_{-},dS_{3}(\Lambda_{eff})\}$ (stable for $\epsilon=-1$; unstable for $\epsilon = 1$) as starting point. In the case ${\epsilon=-1,\lambda_{0}>0}$ these solutions describe accelerating and contracting to Big-Crunch $3D$ Universe, while for the unstable inverse double-well potential $V_{-}$, $i.e.$ in $\{\epsilon=1,\lambda_{0}<0\}$ case, the initial acceleration period ends at the moment $t_a$ ($t_{a}<t_{0}$), when $\tilde{q}(t)$ changes sign and the new deceleration epoch begins. Similarly the eqs. (\ref{23}) when considered in the interval $t_{0}<t<\infty$ represent accelerating and expanding Universe created in the moment $t_{0}$ from the Big-Bang singularity ($R^{(3)}(t_{0})\rightarrow\infty$) towards the infinite future ($t\rightarrow\infty$) $dS_{3}$ - vacuum $\{\sigma_{+},dS_{3}(\Lambda_{eff})\}$ in the ``stable'' $\{\epsilon=-1,\lambda_{0}<0\}$ case. Again the $\{\epsilon=1,\lambda_{0}<0\}$ - Universe exhibits periods of acceleration and deceleration. 

The purpose of this short discussion about the effects of the choice of the i/f c.'s $\{(\sigma_{i,f},\dot{\sigma}_{i,f})\}$ and $\{(\varphi_{i,f},\dot{\varphi}_{i,f})\}$ on the properties of the solutions of eqs. (\ref{quantum}) for the BHT - Matter model (\ref {512}) is to demonstrate the \emph{incompatibility} of the type (2) i/f c.'s, involving scales $L_{2}\le l_{pl}$ (due to $<\tilde{R}^{(3)}>\rightarrow\infty$, $<\dot{\tilde{\varphi}}>\rightarrow\infty$), with the range of validity $L_{2}\gg l_{pl}$ of eqs. (\ref{quantum}). On the other hand, the type (1) i/f. c.'s ($i.e.$ $\sigma^2\le\sigma_{\pm}^2$, $\dot{\varphi}^2<\le4B^2\sigma_{\pm}^2$, etc.) provides an example of ``semiclassical consistency'', when the parameters of the model $\{\epsilon=\pm1, \Lambda_{eff}=M_{gr}^2=\frac{1}{L_{gr}^2}, m_{\sigma}^2=\frac{1}{L_{\sigma}^2}\}$ are chosen such that\footnote{in both cases the treatement of (sub) Planckian scales $L\le l_{pl}$ requires an introduction of other degrees of freedom (than $\hat{h}_{\mu\nu}$) and other methods of quantization } 
\begin{eqnarray*}
l_{pl}\ll L_{gr}<L_{\sigma}
\end{eqnarray*}
(remember $L_{gr}\equiv L_{(1)}$ is the smallest scale in this approximation).

The non-perturbative cosmological solutions of BHT - massive gravity coupled to scalar QFT's with $\sigma^4$ self-interaction (\ref{512}) we have constructed within the restrictions (\ref{m6}), (\ref{galinice2}) imposed by the super-potential method, \emph{do not allow} a complete discussion of the ``unitarity consistency'' of these models. However one can easily verify whether (and for which values of the parameters $\epsilon=\pm1,\lambda_{0}$ and $m_{\sigma}^2$) the above mentioned restrictions are \emph{compatible} with the pure BHT - gravity unitarity conditions \cite{BHT2}:
\begin{eqnarray}
m^2(\Lambda_{eff}-2\epsilon m^2)&>&0\nonumber\\
M_{gr}=-\epsilon m^2+\frac{1}{2}\Lambda_{eff}&\ge&\Lambda_{eff}\label{25}
\end{eqnarray}
together with the well known conditions for unitarity and tachyon-free ``light'' massive scalar $\sigma^4$ - QFT (\ref{lagrange}) in $dS_{3}$ classical background :
\begin{eqnarray}
0<m_{\sigma}^2\le\Lambda_{eff}\label{26}
\end{eqnarray}
We have already realized that the Bounce solutions (\ref{sol}) in both cases $\epsilon\lambda_{0}<0$, $i.e.$ of normal and inverted double well potentials $V_{\pm}(\sigma)$, correspond to the limiting value $\lambda_{BHT}=-\frac{\Lambda}{m^2}=-1$  of the BHT ``cosmological'' parameter $\lambda_{BHT}$. As a consequence the effective cosmological constant
\begin{eqnarray*}
\Lambda_{eff}^{+}=\Lambda_{eff}^{-}=2\vert m^2\vert
\end{eqnarray*}
is equal to the ``graviton $(mass)^2$'', $i.e.$ $M_{gr}^2=\Lambda_{eff}=2\vert m^2\vert$.
In the case: $\epsilon=-1$, $\lambda_{0}>0$ the BHT - u.c.'s (\ref{25}) are automatically satisfied and the restriction (\ref{m6}) now reads
\begin{eqnarray}
4\gamma<m_{\sigma}^2\le2m^2\label{27}
\end{eqnarray}
Therefore the BHT - matter model (\ref{512}) with ``wrong'' sign $\epsilon=-1$ of the curvature term, $\lambda_{0}>0$ and $V_{+}(\sigma)$ representing double well potential (\ref{lagrange}) with $\bar{m}_{\sigma}^2\gamma<0$ and $\gamma>0$  is \emph{unitary} and \emph{tachyons-free}. It corresponds to the lower bound $\lambda_{BHT}=-1$ of the $dS_{3}$ unitary window \cite{BHT2}:
\begin{eqnarray}
-1\le\lambda_{BHT}<0\label{28}
\end{eqnarray}
This special point of the pure BHT - gravity model (\ref{acao}) is known to describe ``\emph{partially massless}'' graviton of one propagating mode, instead of the two helicities $\pm$ 2 graviton massive modes for the generic $M_{gr}^2>\Lambda_{eff}$ case. The observed reduction of the physical degrees of freedom of $\hat{h}_{\mu\nu}$ (for $\lambda_{BHT}=-1$) turns out to be  a result of the enhancement of the gauge symmetries of the linearized (effectively Pauli-Fierz) form of the corresponding equations of motion \cite{des2},\cite{BHT2}. 
 
We have indeed similar non-singular solutions (\ref{sol}) for the case $\epsilon=1$, $\lambda_{0}<0$ ($i.e.$ $m^2<0$) representing the inverted double well potential $V_{-}(\sigma)$ ($\gamma<0$, $\bar{m}_{\sigma}^2>0$). However the BHT - matter model for this range of the parameters is $\it {neither}$ unitary $\it {nor}$ tachyons-free. For $\epsilon=1$, $m^2<0$ and $\Lambda_{eff}=2\vert m^2\vert$ the first of the BHT - u.c.'s (\ref{25}) does not hold and now $\sigma_{\pm}$ - $dS_{3}$ ``vacuum" represent the maxima of $V_{-}(\sigma)$. Hence they are unstable and the corresponding asymptotic ($t\rightarrow\pm\infty$) $\sigma$ - particle states are tachyonic ($m_{\sigma_{\pm}}^2<0$). Due to the ``right'' sign $\epsilon=1$ of the $R$ - term one can try to interprete this model as an effective action for the scalar QFT with $V_{-}(\sigma)$ self-interaction in \emph{classical} gravitational background, thus relaxing the BHT - quantum gravity unitarity conditions (\ref{25}). In order to avoid tachyonic instabilities one should consider ``semiclassical'' states build on the metastable minima of $V_{-}(\sigma)$ ($i.e.$ $\sigma_{0}=0$) and to further study their decay. Note that such state is qualitatively different from the ``constant curvature'' $\vert\sigma_{\pm},dS_{3}(\Lambda_{eff})>$ - states. As one can easily see from eqs. (\ref{sist}), now we have
\begin{eqnarray*}
\dot{\sigma}(0)=2B, \ \ddot{\varphi}(0)=-4B^2, \ \dot{\varphi}(0)=0
\end{eqnarray*}
As a result this relative minima state describes linearly growing matter and (non-constant curvature) Gaussian scalar factor
\begin{eqnarray}
a_{0}^{2}(t)=e^{-2\epsilon B^2 t^2}, \ \sigma_{0}(t)=2Bt\label{ld}
\end{eqnarray}
which makes the analytic study of the spectrum of the fluctuations around it rather complicated.

Let us briefly discuss the remaining two models when $\epsilon\lambda_{0}>0$ (or equivalently $\bar{m}_{\sigma}^2\gamma>0$), $i.e.$
\begin{eqnarray}
V_{\pm}^{cyc}(\sigma)=\pm\frac{\vert\gamma\vert}{4}\Big(\sigma^2+\frac{2\vert m^2\vert}{B^2}\Big)^2-2\Lambda\label{30}
\end{eqnarray}
The solutions of eqs. (\ref{W}) are now periodic  of period $ T = \frac{\pi\vert m\vert}{B^2\sqrt{2}}$ :
\begin{eqnarray}
\sigma_{cyc}(t)=\frac{\vert m\vert\sqrt{2}}{B}\tan\Big(\frac{B^2\sqrt{2}}{\vert m\vert} t\Big) &,&
a_{cyc}^2(t)=\Big\vert\cos\Big(\frac{B^2\sqrt{2}}{\vert m\vert} t\Big)\Big\vert^{\frac{2\epsilon\vert m^2\vert}{B^2}}\label{sol2}
\end{eqnarray}
and  are defined on the finite interval  $t \in \big(-\frac{T}{2},\frac{T}{2}\big)$ only, where   
they represent initial (Big-Bang) and final (Big-Crunch) curvature singularities with $\sigma_{cyc}(\pm\frac{T}{2})\rightarrow\pm\infty$ at both ``ends''. The main problem with these models (and of their solutions (\ref{sol2})) is that they do $\it not$ admit ``vacuum'' solutions of constant curvature. The absolute minima/maxima (for $\epsilon=\pm1$) of $V_{\pm}^{cyc}(\sigma)$, $\sigma_{0}=0$ corresponds to singular non-constant curvature solution (\ref{ld}). Note that the BHT - analysis of the perturbative consistency of $3D$ gravity is based on the linearization $\hat{g}_{\mu\nu}=g_{\mu\nu}^{vac}+\kappa\hat{h}_{\mu\nu}$ around constant curvature stable vacuum states $\vert\sigma_{\pm},g_{\mu\nu}^{vac}>$ ($i.e.$ the ones corresponding to $M_{3}$, $AdS_{3}$, or $dS_{3}$). They are indeed present in the pure BHT - gravity (\ref{acao}) for all values of the parameters $\epsilon$, $\lambda_{0}$ and $\Lambda$. When matter is added, the presence or absence of such type of vacuum solutions depend on the form of the potential $V(\sigma)$ and on the values of the parameters $\epsilon$, $\lambda_{0}$, $m_{\sigma}^2$. As we have seen on the examples of the double-well potentials ($\epsilon\lambda_{0}<0$) the analysis of  perturbative ``semiclassical consistency'' of the BHT - matter models follows the lines of the BHT - arguments \cite{BHT2}, \cite{oda} with special attention to the unitarity and renormalizability of the matter sector. In the $\epsilon\lambda_{0}>0$ type of  models characterized by the potential (\ref{30}) such an analysis requires new methods and their consistency remains an open problem.

We have  avoided up to now the discussion of the important question about the possible manners to couple Matter QFT's to the New Massive Gravity. To begin with the natural questions to ask are: Should one also include higher derivatives  terms like :
$R|\nabla\sigma |^2$ , $|\nabla\sigma |^4$, $R^{\mu\nu}\nabla_{\mu}\sigma\nabla_{\nu}\sigma$,  $RV(\sigma)$ etc.? How such terms might influence the unitarity properties of the corresponding Matter-Gravity model?, etc. Our particular choice (\ref{512}) was motivated by the renormalizability of the matter QFT and we have included in the effective action the UV counter-terms (all of type K for the minimal coupling $\xi=0$) that appear in the lowest order in the couplings $\kappa^2$ and $\gamma$. The analysis of  the structure of divergent diagrams of this graviton -scalar  QFT (in constant curvature background) at orders $\kappa^3$ and $\kappa^4$  suggests that new counter-terms like those mentioned above could be present. This line of arguments however unavoidably leads to the introduction of  large number of new parameters in the corresponding effective action. On the other hand, the original unitarity proof of Bergshoeff, Hohm and Townsend \cite {BHT1} that is based on the equivalence of the NMG action (\ref {acao}) to the specific ``second order''  derivatives (including one more auxiliary spin two massive field $f^{\mu\nu}$) action
\begin{eqnarray}
S_{PF}^{eff}=\frac{1}{\kappa^2}\int dx^3\sqrt{-g}\Big[-R + f^{\mu\nu}G_{\mu\nu} -\frac{m^2}{4}\big(f^{\mu\nu}f_{\mu\nu}-f^2\big)\Big]\label{pf}
\end{eqnarray}
(with $f = g^{\mu\nu}f_{\mu\nu}$) appoints an economic alternative without any new parameters involved. Namely to consider the following rather obvious generalization of the Pauli - Fierz inspired action (\ref {pf}) including matter
fields:
\begin{eqnarray}
S_{PFM}^{eff}=\frac{1}{\kappa^2}\int dx^3\sqrt{-g}\Big\{\epsilon R -\kappa^2\big(\frac{1}{2}|\nabla\sigma |^2+V(\sigma)\big) + f^{\mu\nu}E_{\mu\nu} -\frac{m^2}{4}\big(f^{\mu\nu}f_{\mu\nu}-f^2\big)\Big\}
\label{pfm}
\end{eqnarray}
where we denote
\begin{eqnarray}
 E_{\mu\nu}= G_{\mu\nu} -\frac{\kappa^2}{2}T_{\mu\nu} &,&   
 T_{\mu\nu}(\sigma)=  \nabla_{\mu}\sigma \nabla_{\nu}\sigma -\frac{1}{2}g_{\mu\nu} \nabla_{\rho}\sigma \nabla^{\rho}\sigma-g_{\mu\nu}V(\sigma)
\label{et}
\end{eqnarray}
 The matter potential $V(\sigma)$ has to be chosen by renormalizability arguments. Similarly to the pure gravity case \cite {BHT1}, \cite {des1} one can easily eliminate $f_{\mu\nu}$ in order to get the following suggestive form of the New Massive Gravity - Matter action:
\begin{eqnarray}
S_{NMGM}^{eff}=\frac{1}{\kappa^2}\int dx^3\sqrt{-g}\Big\{\epsilon R +\frac{1}{m^2}K-\kappa^2\Big(\frac{1}{2} | \nabla\sigma |^2+V(\sigma)\Big)\nonumber\\
-\frac{\kappa^2}{m^2}\Big(R^{\mu\nu}T_{\mu\nu}-\frac{1}{4}RT_{\mu}^{\mu}\Big)
+\frac{\kappa^4}{4m^2}\Big(T^{\mu\nu}T_{\mu\nu}-\frac{1}{2}(T^{\mu}_{\mu})^2\Big)\Big\}
\label{nmgm}
\end{eqnarray}
Observe that the last two (new) terms:
\begin{eqnarray}
 U &=& R_{\mu\nu}T^{\mu\nu}-\frac{1}{4}RT_{\mu}^{\mu}= R^{\mu\nu}\nabla_{\mu}\sigma\nabla_{\nu}\sigma - \frac{1}{4}RV-\frac{3}{8}R|\nabla\sigma|^2\nonumber\\
 Y&=& T^{\mu\nu}T_{\mu\nu}-\frac{1}{2}(T_{\mu}^{\mu})^2 = \frac{5}{8}|\nabla\sigma|^4 -\frac{3}{2}V^2- \frac{1}{2}V|\nabla\sigma|^2
\label{uy}
\end{eqnarray}
are of order $\kappa^3$ and $\kappa^4$ correspondingly since the $\kappa$ -expansion of both $R_{\mu\nu}$ and R starts by order $\kappa$ terms. Hence the above action (\ref{nmgm}) when terms up to order $\kappa^2$ only are taken into account  reduces to  the  action (\ref{512}) we have studied. Although the symmetry principles that single out   such higher derivatives Gravity-Matter actions among the many others (with much more free parameters) are not completely clear
their unitarity and renormalizability indeed represent an interesting research problem, which however is out of the scopes of the present paper. Let us  just mention that the unitarity of the matter sector in the complete action (\ref {nmgm}) is rather problematic   due to the presence of the both $|\nabla\sigma|^2$ and $|\nabla\sigma|^4$ terms. The propagator of such massive scalar field contains ghost contributions unless certain fine tunning of the coefficients takes place. 
  
The main objectives of the investigations presented in this paper are the properties and the perturbative consistency of the BHT - matter models (\ref{512}) with specific renormalizable scalar $3D$ QFT added. We have considered as an example the $\sigma ^4$ interaction potential that admits degenerate adiabatic vacuum states. Our goal was the construction of the explicit non-vacuum solutions of the corresponding ``effective" action's equations that possess ``maximal"  $2D$ Poincare symmetry $O(2)\otimes T_2$ and that interpolate between two such vacua. The particular Bounce solutions of the $3D$ Massive BHT - Cosmology with scalar matter (\ref{lagrange}) we have found by superpotential method turns out to saturate the lower bound ${\lambda_{BHT} = -1 }$ ($i.e.$ $\Lambda_{eff}=2|m^2|$) of the BHT unitarity window. There exist strong indications that these solutions are stable as well, due to the fact that they carry $ Z_2 $ topological charges $\pm1$ . The power of the superpotential method we have used in this paper allows the construction of analytic solutions for a vast variety of potentials. The simplest case of linear superpotential we have studied in detail represents just one example selected by the renormalizability of $ \sigma^4 $ interactions in three dimensions. The generalization of our results to the case of simplest periodic superpotential $W(\sigma) = A + B \cos {\sigma} $ that corresponds to the modified Double Sine-Gordon potential $V(\sigma)$ is straightforward. Let us also mention the special case of BHT- gravity coupled to free massless scalar field , $i.e.$ $V(\sigma) = const = -2\Lambda $. In this simple model a family of  unitarity consistent Bounce solutions can be easily found for arbitrary values  of the cosmological constant $ 0 < |\Lambda| < m^2 $ without the use of any superpotentials.
  
Another important line of investigations concerns the construction of domain wall solutions of the Massive 3D Gravity -Matter models (\ref {lagrange}) and their further applications for the reconstruction of the renormalization group flows in the perturbed 2D CFT dual to these BHT-matter models. The well known relation between the FRW cosmological solutions and the static DW's permits an extention of our results to the case  of Janus kink-like DW's for the model (\ref {512}) that interpolate between two stable $AdS_3$ vacua. It turns out that they  correspond  again to the special value $\lambda_{BHT} = -1$, but now with $ \Lambda_{eff} = -2|m^2|$ \cite {prep}.

In conclusion: the unitarity consistency of the simplest Massive 3D Gravity - Matter model (\ref{512}) we have demonstrated on the particular class of cosmological Bounce solutions confirms the expectations that the BHT - model  indeed admits physically interesting renormalizable matter extensions.

$\it {Acknowledgments}$. We are grateful to C.P.Constantinidis for discussions and  for critical reading of the manuscript. This work has been partially supported by PRONEX project
No.35885149/2006 from FAPES-CNPq (Brazil).

\end{document}